\pdfoutput=1
\documentclass[
reprint,
superscriptaddress,
amsmath,amssymb,
aps,
pra,
longbibliography,
]{revtex4-2}
\usepackage{float}
\usepackage{capt-of} 
\usepackage{xcolor}
\usepackage{graphicx}
\usepackage[english]{babel}
\usepackage{dcolumn}
\usepackage{bm}
\usepackage[colorlinks=true, pdfstartview=FitV, linkcolor=blue, citecolor=blue, urlcolor=blue]{hyperref}
\usepackage{dsfont}
\usepackage[separate-uncertainty=true]{siunitx}
\usepackage[version=4]{mhchem}
\graphicspath{{Figures/}}

\begin{document}

\preprint{APS/123-QED}

\title{Determination of the intrinsic mechanical quality factor in high-stress silicon nitride resonators}

\author{Geena Benga}
\affiliation{Laboratory for Solid State Physics, ETH Z\"{u}rich, CH-8093 Z\"urich, Switzerland.}
\affiliation{Quantum Center, ETH Zurich, CH-8093 Zurich, Switzerland}
\author{Vincent Dumont}
\affiliation{Laboratory for Solid State Physics, ETH Z\"{u}rich, CH-8093 Z\"urich, Switzerland.}
\affiliation{Quantum Center, ETH Zurich, CH-8093 Zurich, Switzerland}
\author{Wei Wang}
\affiliation{Laboratory for Solid State Physics, ETH Z\"{u}rich, CH-8093 Z\"urich, Switzerland.}
\author{Nicola Cavalleri}
\affiliation{Institute of Sensor and Actuator Systems, TU Wien, AT-1040 Vienna, Austria.}
\author{Ariane Giesriegl}
\affiliation{Institute of Sensor and Actuator Systems, TU Wien, AT-1040 Vienna, Austria.}
\author{Silvan Schmid}
\affiliation{Institute of Sensor and Actuator Systems, TU Wien, AT-1040 Vienna, Austria.}
\author{Christian L. Degen}
\affiliation{Laboratory for Solid State Physics, ETH Z\"{u}rich, CH-8093 Z\"urich, Switzerland.}
\affiliation{Quantum Center, ETH Zurich, CH-8093 Zurich, Switzerland}
\author{Antonius Armanious}
\email{armanioa@ethz.ch}
\affiliation{Laboratory for Solid State Physics, ETH Z\"{u}rich, CH-8093 Z\"urich, Switzerland.}
\affiliation{Quantum Center, ETH Zurich, CH-8093 Zurich, Switzerland}
\author{Alexander Eichler}
\email{eichlera@ethz.ch}
\affiliation{Laboratory for Solid State Physics, ETH Z\"{u}rich, CH-8093 Z\"urich, Switzerland.}
\affiliation{Quantum Center, ETH Zurich, CH-8093 Zurich, Switzerland}

\date{\today}

\begin{abstract}
Recent advances in silicon nitride nanomechanical resonators have pushed mechanical quality factors to ultra-high values by combining stress-induced dissipation dilution with mode-shape engineering. Neither mechanism alters the intrinsic quality factor $Q_{\mathrm{intr}}$. Targeting the intrinsic loss itself therefore remains an untapped route to even higher $Q$. Doing so first requires reliable quantification of $Q_{\mathrm{intr}}$, which has proven challenging. Here we present a robust methodology that quantifies $Q_{\mathrm{intr}}$ by combining automated mode identification with systematic ringdown measurements over a large number of mechanical modes. Applied to high-stress silicon nitride membranes, it reveals a systematic dependence of $Q_{\mathrm{intr}}$ on thickness that cannot be described using established models, particularly in the ultra-thin limit. We account for this trend with a phenomenological model that incorporates a thickness-dependent loss channel. Together, our method and model open a route toward a microscopic understanding of intrinsic
dissipation and toward directly mitigating its loss channels.

\end{abstract}

\maketitle

\section{Introduction}
Advances in the design and fabrication of ultra-thin silicon nitride (\ce{SiN_x}) resonators have led to substantial improvements in mechanical quality factors. In particular, dissipation dilution enhances the measured quality factor beyond the intrinsic material limit through tensile stress~\cite{bachtold_mesoscopic_2022,fedorov2019generalized,engelsen_ultrahigh-quality-factor_2024}, while mode-shape engineering suppresses clamping and radiative losses by minimizing the mode bending at the anchoring points, resulting in modes with quality factors exceeding $10^9$ at room 
temperature~\cite{tsaturyan_ultracoherent_2017, ghadimi_elastic_2018, catalini_soft-clamped_2020, reetz2019analysis, bereyhi_hierarchical_2022, shin2022spiderweb,cupertino_centimeter-scale_2024}. Low-loss mechanical resonators have promising applications in many fields, including quantum optomechanics~\cite{rossi2018measurement,mason2019continuous,seis2022ground,saarinen_laser_2023}, quantum-limited electro-optical transduction~\cite{bagci2014optical,andrews2014bidirectional,urmey2025high}, photothermal particle detection~\cite{timarac2025nanoplastic,surdu2026quantifying}, integrated optomechanical sensing~\cite{kharbanda2026chip}, scanning force microscopy~\cite{halg_membrane-based_2021, gisler2024enhancing}, spin sensing~\cite{scozzaro2016magnetic,fischer2019spin,karg2020light,kovsata2020spin,thomas2021entanglement,eichler2022ultra,visani2026near}, and topological phononic waveguides~\cite{xi2025soft}.

While dissipation dilution can strongly increase the measured quality factor $Q$ of high-stress mechanical resonators, it does not fundamentally change their intrinsic quality factor, $Q_{\text{intr}}$. Further progress toward enhancing mechanical sensitivity and coherence depends critically on a reliable method for quantifying $Q_{\text{intr}}$. However, this is challenging in high-stress resonators, whose overall damping is typically dominated by clamping and radiative loss. While soft clamping, phononic shielding, and mode shape engineering can strongly suppress these loss terms~\cite{tsaturyan_ultracoherent_2017, ghadimi_elastic_2018, catalini_soft-clamped_2020, reetz2019analysis, bereyhi_hierarchical_2022, shin2022spiderweb, cupertino_centimeter-scale_2024}, such techniques require carefully tailored design and fabrication. They are too involved and costly to be applied routinely to large batches of different samples. There is thus demand for a way to extract $Q_{\text{intr}}$ from resonators with arbitrary design. 

In this work, we report a robust methodology for quantifying $Q_\mathrm{intr}$ in both low- and high-stress mechanical resonators. By measuring the quality factors of a tens to hundreds of mechanical modes with an automated ringdown procedure, we overcome the challenges posed by measurement-to-measurement variation due to the dominant and clamping-dependent radiative loss. In addition, we use a substrate hinge to independently verify the impact of clamping conditions. While a thickness dependence of $Q_\mathrm{intr}$ has previously been observed for low-stress membranes~\cite{villanueva_evidence_2014}, that work used low-stress films, where the larger acoustic mismatch with the substrate suppresses radiation losses and yields less scattered $Q
$ values, allowing $Q_\mathrm{intr}$ to be extracted directly. In high-stress \ce{SiN_x} films, this separation is more difficult and requires full modeling of radiation losses. Our study establishes the trend for devices fabricated and measured under controlled, consistent conditions and, for the first time, extends it to high-stress films. We find a clear monotonic increase of $Q_\mathrm{intr}$ with $h$, consistent across stress regimes, but observe deviations from the standard surface-to-bulk loss model~\cite{yasumura2000quality,villanueva_evidence_2014}, particularly for ultra-thin membranes with $h < \SI{30}{\nano\meter}$. We remedy the discrepancies by incorporating an additional thickness-dependent loss pathway into the phenomenological model. Our model captures the observed behavior well and points toward a microscopic picture of dissipation in nanomechanical resonators.
\begin{figure*}[t]
    \centering
    \includegraphics[width=\linewidth]{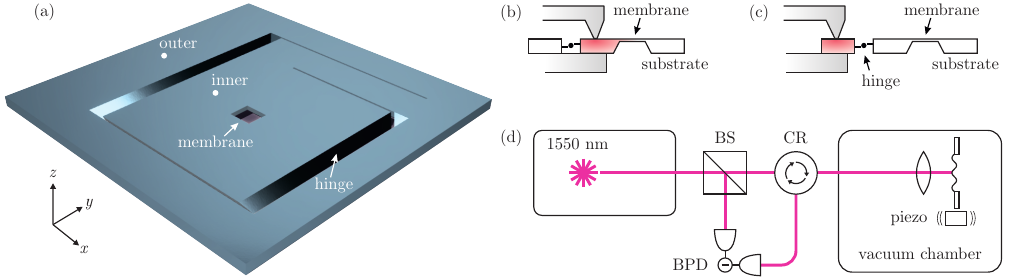} 
    \caption{(a)~To-scale illustration of a \ce{SiN_x} membrane device on the hinged silicon frame. The inner and outer clamping points are highlighted. (b)~Schematic illustration of a device clamped at the inner clamping point. Red shading indicates the assumed model of how stress is induced by the clamping point. (c)~Schematic illustration of a device clamped at the outer clamping point. The hinge protects the membrane from induced stress. (d)~Simplified sketch of the optical Michelson interferometer used for measuring out-of-plane displacement. BS = beam splitter, CR = circulator, BPD = balanced photo diode.}
    \label{fig:Fig1}
\end{figure*}
\section{Mechanical model}
In the high-stress limit, where in-plane tensile stress dominates over bending rigidity~\cite{schmid_fundamentals_2023-1}, square membranes have a series of mechanical out-of-plane modes with resonance frequencies  
\begin{align}\label{eq:modes}
    \omega_i/(2\pi) = \sqrt{\frac{\sigma}{\rho}}\frac{\sqrt{n^2+m^2}}{2L},
\end{align}
where $i=(n,m)$ is the mode index with $n, m \in \mathbb{N}^+$ the numbers of antinodes along $x$ and $y$ direction, respectively, $\omega_i$ is the angular resonance frequency, $L$ the side length of the membrane, $\sigma$ the in-plane tensile stress, and $\rho$ the mass density of \ce{SiN_x}. Each mode can be modeled as a driven, damped linear resonator with the equation of motion:
\begin{align}
    \ddot{z}_i  + \Gamma_i \dot{z}_i + \omega_i^2z_i = F/M_i,
\end{align}
where $z_i$ is the time-dependent out-of-plane displacement of the mode, dots indicate time derivatives, $\Gamma_i$ is the total damping rate, $M_i$ is the effective modal mass, $F = F_0\cos({\omega t})$ is an external driving force with amplitude $F_0$ applied at angular frequency $\omega$, and $t$ is time.

The mode's measured quality factor $Q_i = \omega_i/\Gamma_i$ represents the ratio of the stored to dissipated energy per cycle of oscillation. The total mechanical loss ($Q_{\text{i}}^{-1}$) of a nanomechanical resonator operating in high vacuum, where gas damping is negligible, is given by
\begin{align}
    \frac{1}{Q_i} = \frac{1}{\alpha Q_{\text{rad},i}} + \frac{1}{S_i^{-1}\cdot Q_{\mathrm{intr}}}.
    \label{eq:Qmodel_1}
\end{align}
Here, $Q_{\text{intr}}^{-1}$ denotes the intrinsic loss arising from fundamental material properties, which is commonly separated into bulk and surface loss channels. The factor $S_i = 2\lambda + (n^2 + m^2)\pi^2\lambda^2$ accounts for dissipation dilution, where $\lambda = \frac{h}{L}\sqrt{\frac{E}{12\sigma}}$ is a dimensionless parameter characterizing the flexural bending profile at the membrane boundary depending on $h$, $L$, $\sigma$, and the Young's modulus $E$ of the membrane material ~\cite{schmid_fundamentals_2023-1}. $Q_{\text{rad},i}^{-1}$ originate from the acoustic radiation of phonon energy from the vibrating membrane into the supporting substrate through the physical boundary.

While $Q_{\text{intr}}^{-1}$ is a material property which we assume to be approximately constant for the frequency range spanned by all modes of the resonator, $Q_{\text{rad},i}$ depend on the overlap of, and impedance mismatch between, the membrane and the substrate modes. The analytical calculation of $Q_{\text{rad}}$ from~\cite{wilson-rae_high-q_2011} assumes an `ideal', semi-infinite substrate, where elastic waves radiate away infinitely without ever reflecting back. In a real device, the silicon frame is a finite plate; acoustic energy can reflect off the chip boundaries, and the presence of a clamp to hold the device can introduce additional leakage. The phenomenological scaling parameter $\alpha$ accounts for this. A value of $\alpha=1$ corresponds to the ideal semi-infinite case, while $0<\alpha<1$ capture the increased radiation loss caused by the non-ideal acoustic confinement of the real, finite chip~\cite{wilson-rae_intrinsic_2008}.

Equation~\eqref{eq:Qmodel_1} illustrates the basic issue in isolating $Q_{\text{intr}}$ from measurements. $Q_{\text{rad},i}$ is strongly mode-dependent and cannot be measured directly. Moreover, $Q_{\text{rad},i}$ is sensitive to the precise clamping conditions and can vary substantially between successive mountings of the same chip. Consequently, $Q_{\text{intr}}$ cannot be reliably extracted from a single mode through Eq.~\eqref{eq:Qmodel_1} except if $Q_{\text{rad},i}$ is negligible. To overcome this, we measure up to several hundred mechanical modes of a single membrane and assign mode indices $(n,m)$ by matching the measured frequencies to Eq.~\eqref{eq:modes}. We exploit our knowledge of the measured $Q$ for a large number of modes, together with available analytical expressions for $Q_{\mathrm{rad},i}$ and $S_i$, both mode-dependent, to simultaneously fit $\alpha$ and $Q_{\mathrm{intr}}$, thereby separating radiative and intrinsic
contributions to the measured $Q_i$. 
To assess whether this multi-mode approach is robust against measurement-to-measurement variations in clamping, we repeat the procedure several times on each device for two different clamping geometries. Finally, applying the same protocol to membranes of varying thickness $h$ allows us to separate the volume and surface contributions to $Q_{\text{intr}}$~\cite{villanueva_evidence_2014}.

\begin{figure*}[t]
    \centering
    \includegraphics[width=\textwidth]{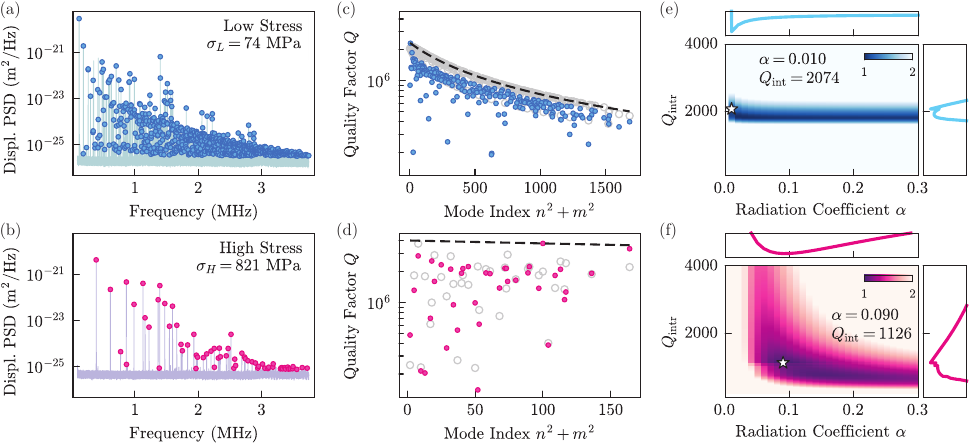}
\caption{Two representative examples of modal analysis and parameter extraction. Upper row: low-stress membrane with $h = \SI{24}{\nano\meter}$ and  $\sigma_L=\SI{74}{\mega\pascal}$. Lower row: high-stress membrane with $h = \SI{23}{\nano\meter}$ and $\sigma_H = \SI{821}{\mega\pascal}$. (a) and (b): thermomechanical displacement Power Spectral Density (PSD) as a function of frequency. Candidate peaks detected by the peak-finding algorithm and successfully matched to a theoretical mode are indicated by circles. (c) and (d): measured quality factors ($Q_i$) plotted against the mode index ($n^2 + m^2$): the experimental results are shown as filled color circles, the theoretical model as empty gray circles, and the diluted intrinsic limit as a dashed line extracted from (e) and (f). (e) and (f): error landscapes generated using the Asymmetric Iteratively Reweighted
Least Square fitting algorithm (see text): the convergence minimum is indicated by a star. The top and right panels display line cuts through the convergence minimum of the relative error ($\mathcal{L}/\mathcal{L}_{\mathrm{min}}$) in linear scale. 
}
    \label{fig:Fig2}
\end{figure*}
\section{Device and Experimental Setup}\label{sec:Dev_and_Setup}
Our devices are $\SI{1}{\milli\meter}\times \SI{1}{\milli\meter}$ \ce{SiN_x} membranes, fabricated by Qfactory (DK), supported by rigid silicon wafers, see Fig.~\ref{fig:Fig1}(a). We use both stoichiometric high-stress and Si-rich low-stress films (both referred to as \ce{SiN_x}) with thicknesses $h$ ranging from \SI{20}{\nano\meter} to \SI{240}{\nano\meter}. To test the effect of clamping conditions~\cite{rieger2014energy}, we engineer the frame to include grooves that decouple the clamping point from the membrane, referred to as hinged frame, see Fig.~\ref{fig:Fig1}(b) and (c). For every experiment, we measure a membrane three times in each of the two configurations: first with `inner clamping', where the chip is fixed directly at the inner frame, and then with `outer clamping', where the chip is fixed at the outer perimeter, with the hinge intended to isolate the membrane from the clamping point.

Our devices are mounted inside a high-vacuum chamber (pressure $p \approx \SI{e-7}{\milli\bar}$) to eliminate viscous damping. Mechanical oscillations are driven with a piezoelectric element. The out-of-plane displacement is measured using an optical Michelson interferometer: a $\SI{1550}{\nano\meter}$ laser is directed through a beam splitter and focused onto the center of the membrane using a GRIN lens. The light reflected from the membrane surface interferes with the light reflected from the GRIN lens and is directed to a balanced photodetector, as sketched in Fig.~\ref{fig:Fig1}(d).

\section{Measurement procedure and analysis}
Our procedure to determine $Q_\mathrm{intr}$ relies on measuring $Q_i$ for a large number of mechanical modes of a membrane device, see Fig.~\ref{fig:Fig2}. We develop an automated process comprising three steps:

(i)~\emph{Mode detection and cross-validation.} We measure the thermomechanical displacement noise power spectral density (PSD) of the membrane and apply a peak-finding algorithm, see Fig.~\ref{fig:Fig2}(a) and (b). Detected peaks are cross-referenced with Eq.~\eqref{eq:modes}, and only those matching a predicted $\omega_i$ within a defined tolerance are retained. See~\ref{sec:modeid} for details.

(ii)~\emph{Ringdown measurement of $Q_i$.} We drive each identified mode resonantly using the phase-locked loop of lock-in amplifier (MFLI, Z\"urich Instruments). When switching the drive off, we obtain the free decay of the mechanical mode. We then fit the free decay to an exponential to extract $Q_i$, see~\ref{sec:RDallmodes}.  Curves with low signal-to-noise ratio, beating patterns indicative of mode coupling~\cite{groblacher_substrate_mode_coupling, norte_beating_ringdowns}, or non-exponential decay are rejected, see~\ref{sec:criteria}. 

(iii)~\emph{Fitting Eq.~\eqref{eq:Qmodel_1} to our measured $Q_i$ values.} We assume that $Q_\mathrm{intr} S_i^{-1}$ sets an upper bound on $Q_i$, with deviations to lower values attributed to $Q_{\mathrm{rad},i}$. This bound is clearly discernible in low-stress membranes, see Fig.~\ref{fig:Fig2}(c), but obscured in high-stress membranes where $Q_{\mathrm{rad},i}$ dominates, as shown in Fig.~\ref{fig:Fig2}(d). Standard least-squares fits fail to include rigid boundaries for the fit parameters, and instead assume symmetric cost functions around a parabolic minimum (best fit). We therefore implement an Asymmetric Iteratively Reweighted Least Squares (Asymmetric IRALS)~\cite{AsymIRALS} algorithm by defining a cost function in $(Q_\mathrm{intr},\alpha)$ and constraining the allowed values of the parameters, see~\ref{sec:IRALS} and Fig.~\ref{fig:Fig2}(e) and (f).

\begin{figure*}[t]
    \centering
    \includegraphics[width=\textwidth]{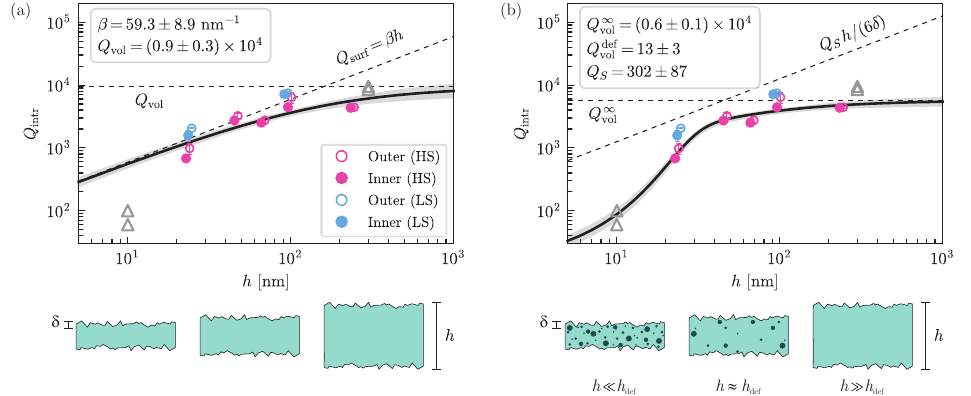}
    \caption{
    Thickness dependence of the intrinsic quality factor $Q_{\text{intr}}$. Pink and blue data markers correspond to high and low stress devices, respectively, and are shown with a small offset in $h$ for better visibility. 
    Each marker is the mean of three independent measurement under identical conditions, with error bars representing the standard deviation. Inner and outer clamping points are slightly offset in $h$ for clarity.
    See legend in (a) for symbol meaning. Solid curves represent model fits to the high-stress data using (a) Eq.~\eqref{eq:loss_model} and (b) Eq.~\eqref{eq:composite_model}, with their associated $\pm1\sigma$ uncertainty bands shaded in gray. Dashed lines are asymptotic limits. Gray triangles are data points from a different membrane batch and measured with a different setup. They are used for independent verification only. Schematic illustrations under the panels visualize the model dependence on $h$: in Eq.~\eqref{eq:loss_model}, changes in $h$ do not affect the bulk properties. In Eq.~\eqref{eq:composite_model}, we assume a defect density that modifies the bulk loss and that exponentially depends on $h$ with a characteristic length thickness $h_\mathrm{def}$. For $h\gg h_\mathrm{def}$, the defect density becomes negligible.
    }
    \label{fig:Fig3}
\end{figure*}

\section{Results and discussion}
Figure~\ref{fig:Fig3}(a) shows the results of the fit parameter $Q_\mathrm{intr}$ as a function of $h$ for two low-stress (LS) and five high-stress (HS) membranes, measured in both inner and outer clamping configurations. Each point is the mean of three independent measurement and fitting cycles. 

Our central observation in Fig.~\ref{fig:Fig3}(a) is a monotonic increase of $Q_\mathrm{intr}$ with $h$ in both stress regimes. Such a dependence has previously been reported for low-stress \ce{SiN_x} resonators from a meta-analysis of literature data~\cite{villanueva_evidence_2014}, but extending it to high-stress films remained elusive: there, $Q_{\mathrm{rad},i}$ dominates and is strongly mode-dependent, requiring the two loss channels to be decoupled.

The values of $Q_{\mathrm{intr}}$ extracted with outer clamping have a consistent offset from those with inner clamping, but the differences are not significant compared to the changes as a function of $h$. We conclude that our approach yields a sufficiently accurate quantification of $Q_{\mathrm{intr}}$ to serve as a basis for investigating the underlying mechanisms. The extracted radiation coupling parameter $\alpha$ is shown in the Appendix~\ref{sec:alpha}. A full understanding thereof remains an open question and a direction for future work.

Next, we use the phenomenological model of Ref.~\cite{yasumura2000quality} to explain $Q_\mathrm{intr}(h)$ by a combination of volume loss and surface-defect loss, 
\begin{align}
\frac{1}{Q_{\text{intr}}(h)} = \frac{1}{Q_{\text{vol}}} + \frac{1}{Q_{\text{surf}}(h)} .
\label{eq:loss_model}
\end{align}
$Q_{\text{vol}}$ is an intrinsic material property and independent of $h$. $Q_{\text{surf}}(h) = \beta h$, with $\beta$ a constant coefficient, captures losses from a thin defect-rich surface layer. Its contribution is diluted as the membrane thickness increases, since the surface-to-volume ratio scales as $1/h$. 
Fitting Eq.~\eqref{eq:loss_model} to our high-stress data in Fig.~\ref{fig:Fig3}(a) yields $\beta = 59 \pm 9$\,nm$^{-1}$ and $Q_\mathrm{vol} = (0.9 \pm 0.3)\times 10^4$. Both values match the previously measured value for low-stress membranes within a small margin~\cite{villanueva_evidence_2014}.  
We find that the two-parameter model of Eq.~\eqref{eq:loss_model} captures the broad trend of $Q_\mathrm{intr}(h)$ well at intermediate and large thicknesses, but we note in our data a stronger curvature towards lower $Q_{\text{intr}}$ for small $h$, see Fig.~\ref{fig:Fig3}(a). This mismatch indicates the presence of an additional loss channel that becomes relevant for $h \lesssim \SI{30}{\nano\meter}$ and is not captured by a simple combination of volume and surface losses.

To account for this additional loss mechanism, we construct a model that extends Eq.~\eqref{eq:loss_model} with a further thickness-dependent excess volume loss term:
\begin{align}
    \frac{1}{Q_\mathrm{intr}(h)} =
    \frac{1}{Q_\mathrm{vol}^\infty}
    + \frac{6\delta}{h}\,\frac{1}{Q_\mathrm{S}}
    + \frac{1}{Q_\mathrm{vol}^{\mathrm{def}}}\,e^{-h/h_\mathrm{def}}.
    \label{eq:composite_model}
\end{align}
The first term, $1/Q_\mathrm{vol}^\infty$, corresponds to the first term in
Eq.~\eqref{eq:loss_model} and represents the $h$-independent losses. The second term describes a tensile-stress-dominated plate with two surface layers, each with a loss $1/Q_\mathrm{S}$ and thickness $\delta$, sandwiching a bulk region. The value of $\delta=\SI{0.4}{\nano\meter}$ is measured from the surface roughness of LPCVD \ce{SiN} films (see~\ref{sec:delta}). As derived in~\ref{sec:loss_derivation}, the surface-related loss in the limit $h \gg \delta$ is given by $\frac{6\delta}{h}\,\frac{1}{Q_\mathrm{S}}$ (same as in Eq.~\eqref{eq:loss_model}, where $\frac{1}{Q_{\text{surf}}(h)} =\frac{1}{\beta h} = \frac{6\delta}{h} \frac{1}{Q_\mathrm{S}}$) \cite{yasumura2000quality}. Finally, the third term is an additional thickness-dependent loss contribution that corrects for the underestimation of the losses by Eq.~\eqref{eq:loss_model} in ultra-thin membranes. We parameterize it as an exponential correction,
$\frac{1}{Q_\mathrm{vol}^{\mathrm{def}}}\,e^{-h/h_\mathrm{def}}$, that adds to the asymptotic bulk loss $\frac{1}{Q_\mathrm{vol}^\infty}$. Here, $Q_\mathrm{vol}^{\mathrm{def}}$ corresponds to an excess loss associated with the characteristic length scale $h_\mathrm{def}$, of unknown origin. We use the subscript `$\mathrm{vol}$' to indicate that this loss may originate anywhere within and on the membrane, and that it could potentially share a common scaling with $\frac{1}{Q_\mathrm{vol}^\infty}$. 

The microscopic origin of the exponential defect-layer term cannot be determined from our measurements alone. It may reflect a population of defects whose effect on dissipation grows as the film thins, or a structurally distinct ultrathin-film regime arising from the LPCVD growth process itself. The latter possibility finds support in independent observations from the thin-film literature: \ce{SiN_x} growth on native silicon oxide is known to proceed via island nucleation followed by coalescence~\cite{copel_nucleation_1999}. The early stages of growth are retarded, which is attributed to slow nucleation~\cite{martin_1991_retardation}. Moreover, the electrical integrity of LPCVD SiN films breaks down for membranes of thickness below $\sim\SI{5}{\nano\meter}$, with sharply increased leakage currents and pinhole formation~\cite{weinberg_1990}. Taken together, these results suggest that films thinner than the island-coalescence length are structurally incomplete and defect-rich. Adopting this length scale as a starting point, we fix $h_\mathrm{def} = \SI{5}{\nano\meter}$.
With $\delta$ and $h_\mathrm{def}$ fixed, fitting Eq.~\eqref{eq:composite_model} to the high-stress dataset yields $Q_\mathrm{vol}^\infty \approx 6000$, $Q_\mathrm{vol}^{\mathrm{def}}\approx 13$ and $Q_S \approx 302$, see Fig.~\ref{fig:Fig3}(b). 

As an independent test, we add to the plot $Q_\mathrm{intr}$ values from high-stress membranes at $h = \SI{10}{\nano\meter}$ and $h = \SI{300}{\nano\meter}$, fabricated and measured independently of this work (grey triangles). These points agree well with Eq.~\eqref{eq:composite_model} at both extremes of the thickness range, even though they are not used for the fit. In particular, the \SI{10}{\nano\meter} data lie far below the prediction of the simpler Eq.~\eqref{eq:loss_model} but are captured by Eq.~\eqref{eq:composite_model}, providing direct support for the inclusion of the exponential defect-layer term.

\section{Summary and Outlook}
\label{sec:outlook}

We demonstrate a robust methodology for quantifying the intrinsic quality factor $Q_\mathrm{intr}$ of high-stress \ce{SiN_x} membranes, particularly for thin membranes $h< \SI{100}{\nano\meter}$. Our results confirm that high-stress resonators have a similar dependency of $Q_\mathrm{intr}$ on thickness $h$ as their low-stress counterparts. However, we find that the previous model for this dependency, comprising two terms to account for volume and surface loss, is insufficient in the regime of very thin membranes. A generalization of the model with an additional thickness-dependent loss channel improves the modeling considerably. The exact microscopic origin, however, remains an open question for future work: whether the loss originates at the free surface, at the growth interface, or is distributed throughout the bulk remains to be determined.

We show that there is a strong motivation to better understand radiative and clamping loss to improve the accuracy of determining $Q_\mathrm{intr}$ in high-stress devices. The model in Ref.~\cite{wilson-rae_high-q_2011} treats the substrate as a semi-infinite plane, an approximation which does not hold for small chips. We believe this to be one of the main source of uncertainty in our methodology.

The methodology developed here provides a reliable experimental handle on $Q_\mathrm{intr}$, opening the door to systematic studies of how growth conditions, surface treatments, and post-processing steps affect intrinsic dissipation, and thereby a path toward engineering lower-loss \ce{SiN_x} films. Furthermore, the model in Eq.~\eqref{eq:composite_model} points to a potential microscopic source for dissipation, which will be investigated in more detail in consequent studies. Ultimately, this will lead to improving the performance of nanomechanical sensors beyond the state of the art~\cite{eichler2022ultra}.

\emph{Acknowledgments.} We thank S. Misra, L. Martins Mestre, N. Prumbaum, D. Visani, K. Knapp and B. Kharbhanda for the fruitful discussions. This research was possible thanks to the Swiss National Science Foundation (SNSF) through Grants No. 200021E\_209235, the Novo Nordisk Foundation through Grant No. NNF22OC0077964, and the Austrian Science Fund (FWF) 10.55776/I6086. V.D. acknowledges support from the ETH Zurich Postdoctoral Fellowship Grant No. 23-1 FEL-023 and the Swiss National Science Foundation (SNSF) Postdoctoral Fellowship Grant 217118. 
 
\bibliography{bibl}

\clearpage
\onecolumngrid
\begin{center}
  \textbf{\large Supplementary Information}
\end{center}
\vspace{1cm}

\setcounter{section}{0}
\setcounter{figure}{0}
\setcounter{equation}{0}
\setcounter{table}{0}
\renewcommand{\thesection}{S\arabic{section}}
\renewcommand{\thefigure}{S\arabic{figure}}
\renewcommand{\theequation}{S\arabic{equation}}
\renewcommand{\thetable}{S\arabic{table}}
\section{Automated Ringdown Methodology}
\label{sec:autoRD}
\subsection{Mode Identification}
\label{sec:modeid}
Prior to performing ringdown measurements, an automated peak-detection algorithm was used to identify the resonance frequencies of the fundamental and higher-order mechanical modes. First, the thermal noise power spectral density (PSD) of the membrane was acquired. A peak-finding algorithm (\texttt{scipy.signal.find\_peaks}) was applied to the PSD to locate local maxima, using a prominence threshold and a minimum peak distance.

To ensure that only valid square membrane modes were characterized, the detected peaks were cross-referenced with the theoretical dispersion relation for a square membrane under high tensile stress:
\begin{align}
    f_{nm} = f_{1,1} \sqrt{\frac{n^2 + m^2}{2}}
\end{align}
where $f_{1,1}$ is the frequency of the fundamental mode, and $n, m$ are the mode indices. Only detected peaks that matched a predicted theoretical frequency within a defined tolerance window (e.g., $\pm \SI{10}{\kilo\hertz}$) were selected for subsequent ringdown measurements. This filtering step prevented the system from locking onto spurious peaks or mixing products.

\subsection{Ringdown Acquisition}
\label{sec:RDallmodes}
To handle the large number of mechanical modes identified in the previous step, we used an automated Python control script to interface with the MFLI lock-in amplifier. The data acquisition procedure for each target mode follows a rigorous logic:

\begin{enumerate}
    \item \textbf{Initialization:} The specific mode frequency is targeted, and the drive amplitude is initialized.
    \item \textbf{Signal Optimization:} The script monitors the response amplitude. If the signal is below a threshold and the drive is not saturated, the drive voltage is doubled iteratively. A noise check is performed; if doubling the drive does not result in a proportional signal increase (ratio $< 1.1$), the mode is flagged as noise and skipped.
    \item \textbf{Phase-Locked Loop (PLL):} To ensure the ringdown starts from the maximum amplitude, a PLL is engaged. The system performs a rough phase sweep (steps of \SI{90}{\degree}) followed by a fine sweep to determine the optimal phase setpoint. The PLL locks to the resonance frequency, and if the phase error remains high, the drive is increased further until a stable lock is achieved.
    \item \textbf{Ringdown Acquisition:} Once locked, the drive is disabled, and the ringdown decay is recorded.
\end{enumerate}

\subsection{Data Selection Criteria}
\label{sec:criteria}

\begin{figure}[h]
    \centering
    \includegraphics[width=\linewidth]{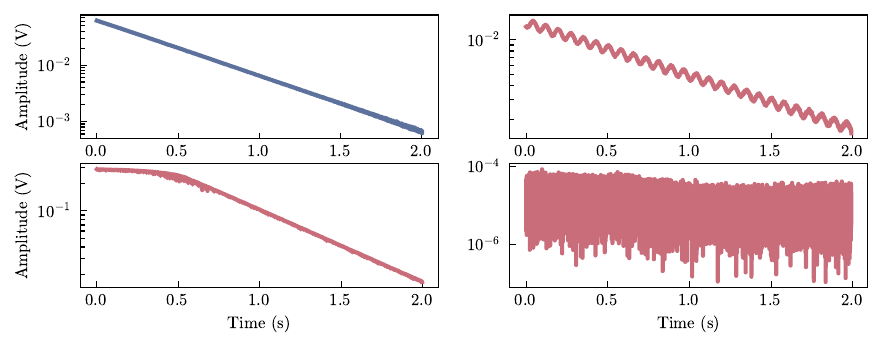} 
    \caption{Criteria for valid Q-factor determination. Upper left: An accepted exponential decay. Rejected cases include mode beating (upper right), non-linear amplitude dependence (lower left), and failed lock or high noise floor (lower right).}
    \label{fig:s_criteria}
\end{figure}

To ensure the validity of $Q_{\mathrm{intr}}$ extraction, raw ringdown curves undergo a strict selection process. We classify measurements into "Accepted" and "Rejected" categories based on the linearity of the log-magnitude decay and the fit residuals. Figure~\ref{fig:s_criteria} illustrates the classification criteria:
\begin{itemize}
    \item \textbf{Accepted:} A clean, single-exponential decay appearing linear on a logarithmic scale.
    \item \textbf{Rejected (Beating):} Evidence of strong amplitude modulation (beating), indicating coupling to a nearby mechanical mode.
    \item \textbf{Rejected (Non-Linear):} Decay profiles exhibiting non-linear damping characteristics (e.g., Duffing nonlinearity), where the decay rate changes with amplitude.
    \item \textbf{Rejected (No Lock):} Cases where the PLL failed to lock onto the resonance or the signal-to-noise ratio was insufficient, resulting in a noise-dominated trace.
\end{itemize}

\section{IRALS Extraction of \texorpdfstring{$Q_\mathrm{intr}$}{Qintr} and $\alpha$}
\label{sec:IRALS}
A key complication in extracting $Q_\mathrm{intr}$ is that different mechanical modes of the same membrane experience different radiative losses depending on their spatial overlap with the boundary geometry. To separate $Q_\mathrm{intr}$ from $Q_\mathrm{rad}$, we fit the full modal dataset using the Wilson--Rae model, incorporating numerically simulated radiation quality factors $Q_{\mathrm{rad},i}$. The model predicts a total quality factor governed by:

\begin{equation}
    \frac{1}{Q_\mathrm{model,i}} = \frac{S_i}{Q_\mathrm{intr}} + \frac{1}{\alpha \, Q_{\mathrm{rad},i}},
    \label{eq:Qmodelsupp}
\end{equation}

where $S_i$ is the normalized mode-shape strain integral and $\alpha$ is a clamping coupling scaling parameter.
The IRALS algorithm performs a 2D grid search over $(\alpha, Q_\mathrm{intr})$, minimizing the weighted log-error

\begin{equation}
    \mathcal{L}(\alpha, Q_\mathrm{intr}) = \frac{1}{N}\sum_{i=1}^{N} w_i
    \left(\log (Q_{\mathrm{meas},i}) - \log (Q_{\mathrm{model},i})\right)^2,
\end{equation}

where per-mode weights $w_i$ are updated iteratively to enforce the envelope constraint. A penalty is applied only when $Q_{\mathrm{meas},i} >
Q_\mathrm{intr}/S_i$; once the model envelopes a mode, its weight freezes. Convergence is reached when the theoretical curve is a complete upper bound to all data points.
Because the error landscape can develop a one-sided plateau along the $Q_\mathrm{intr}$ axis — arising when radiative losses dominate and
$1/Q_\mathrm{intr}$ becomes negligible — we use a threshold-based valley-entry condition rather than the absolute minimum to extract parameters. The radiation coefficient $\alpha$ is fixed at the landscape minimum; $Q_\mathrm{intr}$ is then set to the lowest value whose error falls within a defined entry tolerance (0.8\% above the minimum). This conservatively lower-bounds $Q_\mathrm{intr}$, ensuring that residual radiative contributions do not inflate the extracted value. Asymmetric confidence bounds are derived from 1D slices of the normalized error landscape: the lower bound is sharply defined, whereas the upper bound can extend to infinity when the plateau is entered, faithfully reflecting the physical insensitivity of the data in that regime.
\newpage
\subsection{Clamping parameter $\alpha$}
\label{sec:alpha}
\begin{figure}[h]
    \centering  
    \includegraphics[]{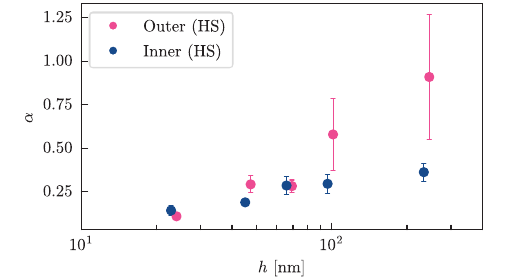} 
    \caption{Thickness dependence of the radiation coupling parameter $\alpha$ for high-stress membranes. The extracted $\alpha$ values are shown for both outer (pink circles) and inner (blue circles) clamping configurations and are shown with a small offset in $h$ for better visibility.}
    \label{fig:Fig4}
\end{figure}

The behavior of the clamping parameter $\alpha$ is shown in Fig.~\ref{fig:Fig4}: for inner clamping, $\alpha$ shows only a weak dependence on $h$, while for outer clamping the increase with $h$ is considerably stronger, which would be consistent with the hinge progressively mitigating phonon leakage as the natural impedance mismatch between membrane and substrate decreases. The two configurations converge at the thinnest membranes, where the large impedance mismatch renders the hinge ineffective.
The observed linear scaling of $\alpha$ with $h$ for both clamping conditions was unexpected: within the semi-infinite substrate model, $\alpha$ should be a geometry-dependent constant independent of $h$. We tentatively attribute this dependency to the finite thickness of the Si frame, which may become comparable to the wavelength of the radiated elastic waves, violating the semi-infinite assumption.
\subsection{$Q_\mathrm{intr}$ Model}
\subsubsection{Surface Roughness AFM Measurement}
\label{sec:delta}
\noindent
\begin{minipage}[c]{0.4\textwidth}
The surface topology was analyzed using ScanAsyst equipment (Bruker, Dimension Edge with Scan Asyst).  An antimony-doped silicon tip (Value AFM Probes by Bruker, $T = 4\ \mu m$, $k = 42\ N/m$, $f_0 = 320\ kHz$) with reflective aluminium on the back side was used in tapping mode. Each measurement was done on the frame of the sample, which was prior cleaned with isopropanol and gentle \ce{N2} stream. Images were recorded using height, phase and amplitude channels of square frames of $\SI{1}{\micro\meter}\cdot\SI{1}{\micro\meter}$ with a resolution of \SI{512}{\nano\meter} in both x and y direction and a scanning rate of $\SI{0.5}{\micro\meter\per\second}$. The root-mean-squared (RMS) surface roughness was evaluated with the help of the python surfalize package~\cite{Schell2024}, with the results for several stress regimes listed in Tab.~\ref{tab:rms_roughness}.

\vspace{1em} 
\centering
\begin{tabular}{|l|c|c|c|c|}
    \hline
    & Silicon & 100 MPa & 200 MPa & 1.1 GPa \\
    \hline
    RMS [pm] & 63 & 536 & 448 & 433 \\
    \hline
\end{tabular}
\captionof{table}{RMS roughness measurements}
\label{tab:rms_roughness}
\end{minipage}
\hfill 
\begin{minipage}[c]{0.55\textwidth}
    \centering  
    \includegraphics[width=\linewidth]{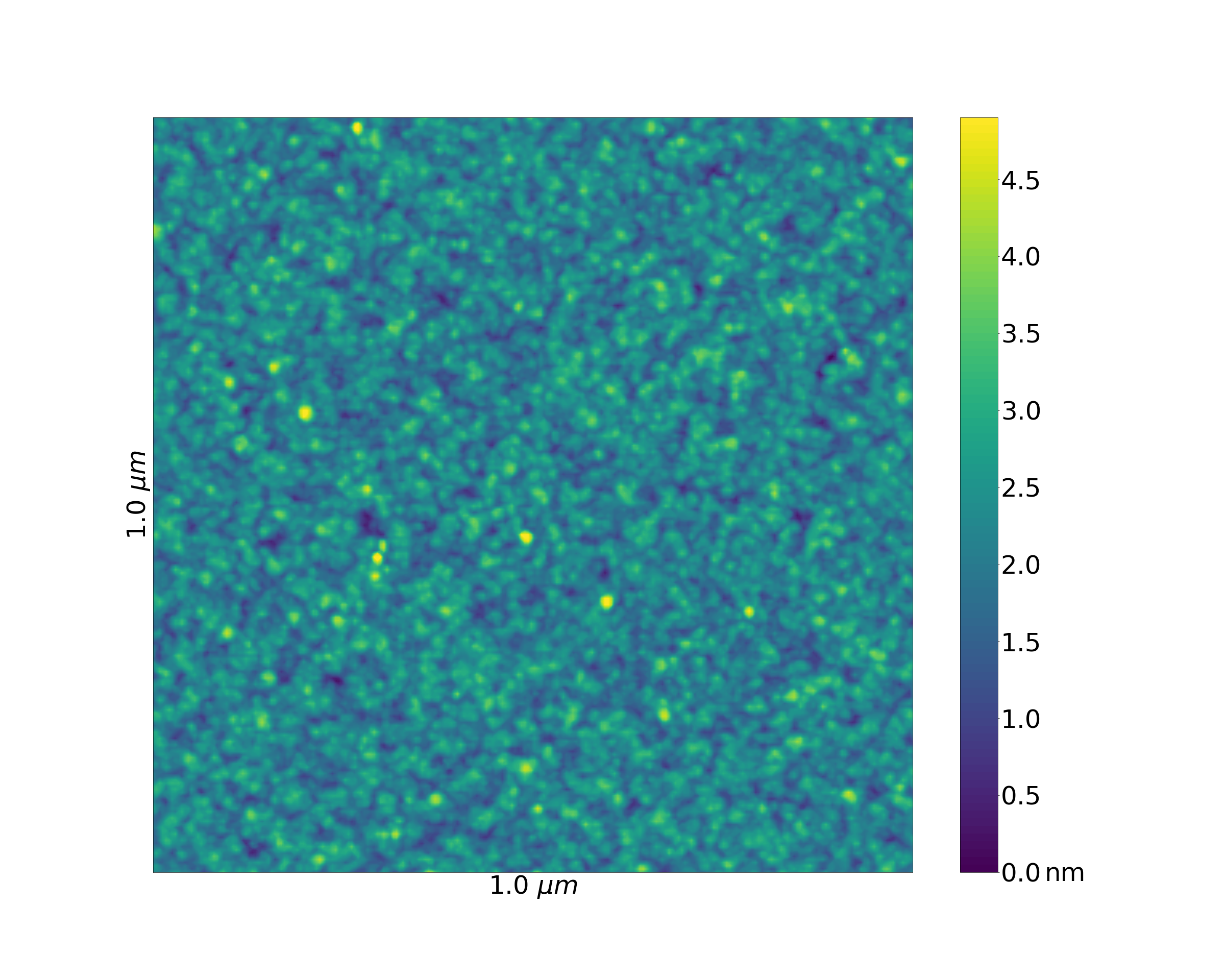} 
    \captionof{figure}{Example AFM scan of the $\SI{100}{\mega\pascal}$ device.}
    \label{fig:AFM}
\end{minipage}

\subsubsection{Composite Model}
\label{sec:loss_derivation}

Applying Kirchhoff bending kinematics — in-plane strain $\varepsilon \propto z$, hence $z^2$-weighted bending energy — to a plate with a lossy surface layer of thickness $\delta$ yields
\begin{equation}
    \frac{1}{Q(h)} =
    \frac{(h-2\delta)^3}{h^3}\,\frac{1}{Q_\mathrm{vol}(h)}
    +
    \frac{h^3-(h-2\delta)^3}{h^3}\,\frac{1}{Q_S}.
    \label{eq:composite}
\end{equation}
In the thick limit $\delta \ll h$ this recovers $Q \approx Q_S h/(6\delta)$, the Yasumura linear asymptote, with $\beta = Q_S/(6\delta)$. In the thin limit $h \to 2\delta$, the surface layer fills the cross-section and $Q$ saturates at $Q_S$, avoiding the unphysical extrapolation to zero. It is important to note, however, that this cubic saturation acts as a \emph{positive}  correction: it raises $Q$ above the linear asymptote at small $h$, precisely the opposite direction of our observations. The sub-asymptotic dip therefore requires an additional loss mechanism. We model this as a thickness-dependent excess bulk loss whose loss tangent decays exponentially into the film with a characteristic length $h_\mathrm{def}$:
\begin{equation}
    \frac{1}{Q_\mathrm{vol}(h)} = \frac{1}{Q_\mathrm{vol}^\infty} + \frac{1}{Q_\mathrm{vol}^{{\mathrm{def}}}}e^{-h/h_\mathrm{def}},
    \label{eq:TLS}
\end{equation}
with $Q_\mathrm{bulk}^\infty$ the asymptotic bulk quality factor and $\frac{1}{Q_\mathrm{bulk}^{\mathrm{def}}}$ the excess loss amplitude at the interface. Substituting into \ref{eq:composite} gives the full combined model:
\begin{equation}
    \frac{1}{Q(h)} = \frac{h^3-(h-2\delta)^3}{h^3}\cdot\frac{1}{Q_S}+
    \frac{(h-2\delta)^3}{h^3}\cdot\left( \frac{1}{Q_\mathrm{vol}^\infty} + \frac{1}{Q_\mathrm{vol}^{{\mathrm{def}}}}e^{-h/h_\mathrm{def}}\right).
    \label{eq:combined}
\end{equation}
 with $\delta = 0.4$\,nm (surface roughness scale), and $h_\mathrm{def} = 5$\,nm (from electrical leakage data on ultrathin LPCVD SiN, which shows a strong increase in leakage currents below 5--6\,nm~\cite{kwakman_1993,weinberg_1990}).

\end{document}